# Role of NGOs in fostering equity and social inclusion in cities of Bangladesh: The Case of Dhaka

Hossain Ahmed Taufiq[1]

**Abstract:** Non-governmental organisations have made a significant contribution in the development of Bangladesh. Today, Bangladesh has more than 2000 NGOs, and few of them are among the largest in the world. NGOs are claimed to have impacts on the sustainable development in Bangladesh. However, to what extent they have fostered equity and social inclusion in the urban cities of Bangladesh remains a subject of thorough examination. The 11th goal of the Sustainable Development Goals (SDG) advocates for making cities and human settlements inclusive, safe, resilient and sustainable. Bangladesh which is the most densely populated country in the world faces multifaceted urbanisation challenges. The capital city Dhaka itself has experienced staggering population growth in last few decades. Today, Dhaka has become one of the fastest growing megacities in the world. Dhaka started its journey with a manageable population of 2.2 million in 1975 which now reached 14.54 million. The growth rate averaged 6 per cent each year. As this rapid growth of Dhaka City is not commensurate with its industrial development, a significant portion of its population is living in informal settlements or slums where they experience the highest level of poverty and vulnerability. Many NGOs have taken either concerted or individual efforts to address socio-economic challenges in the city. Earlier results suggest that programs undertaken by NGOs have shown potential to positively contribute to fostering equity and reducing social exclusion. This paper, attempts to explore what types of relevant NGO programs are currently in place taking the case of Dhaka city.

---

[1] Hossain Ahmed Taufiq, Lecturer, Global Studies and Governance Program, Independent University, Bangladesh (IUB). Email: taufiqh@tcd.ie

# INTRODUCTION

Dhaka, the capital of Bangladesh, has grown rapidly in last few decades, and this fast growth is expected to continue. It means, Dhaka will remain the dominant urban agglomeration of the country and will continue to accommodate a greater portion of new urban population than other major oldest cities. One significant reason is migration. Studies reveal that 84.6% population of Dhaka is migrant and the majority of migration is rural to urban in nature (85 percent), which is followed by urban to urban migration (seven percent) (Rahman 2011; PPRC 2009). Dhaka became the subject of migration for numerous reasons. Search for better employment and better business opportunities play a vital role in such migrations. About 70% Dhaka's population stated that their jobs are the reason, while 20% cited better business opportunities in the capital (PPRC, 2009), while about 23% population moved to the major metropolitan city for their children's education (Rahman, 2011). According to Riaz (2016) 15.3% of the population in Dhaka and Chittagong are students, and they cited that Dhaka has the best education facilities followed by Chittagong. Hope for better living standards, charm for colourful metropolitan life and access to better healthcare facilities are the other contributory factors (Jahan, 2012).

Large inbound migrations are supposed to increase productivity in the city areas (Haworth, Long, & David, 1978; Au & J. Vernon, 2006). Despite having decades of massive inbound migration, the urban economic growth of Dhaka failed to commensurate with it, and therefore urban poverty persists (Hossain 2010). Lack of adequate and affordable housing in Dhaka forced a significant portion of the migrant population to live in squatter settlements or slums where they experience the highest level of poverty and vulnerability. Although the roots of slum and squatter settlements in Dhaka are as old as the city itself, the prolific growth of these informal settlements took place right after the 1971 independence (Hossain 2008). In the following years of independence, Dhaka saw many people started living in these informal settlements. There are three probable reasons behind such proliferation. First, shortage of adequate housing due to heavy loss and damage in the independence war. Second, in a newly independent war-ravaged country, suddenly many people saw that they no longer possess the financial means to avail decent housing anymore. Third, Bangladesh suffered a deadly famine in 1974, leading to mass migration in the urban areas (Razzaque, 1985). Nevertheless, in 1975, attempts/ approaches have been undertaken by the then national government and local authorities to address the slum/ squatter situations in the Dhaka city (Mohit, 2012). These led to a mass eviction of slums and subsequent rehabilitation of the dwellers in other locations and facilities (Rashid, 2009). These significantly reduced the number of dwellers and slums in Dhaka at that time. In 1976, only 10 slums existed with a population totalling only 10,000 (Hossain, 2010). In 1993, it increased to 2,156 settlements with a population of 718,143 and in 1996 it increased to 3,007 settlements with a population of 1.1 million (CUS 1993; 1996). Most of the slum settlements mushroomed in the last three decades but the highest growth was experienced between 1981-1990 (Jahan, 2012). A study based on Dhaka (Ahmed 2007) shows that around 2.84 million people live in 4,000 slums and squatters accounting for 30 percent of the city's people (Jahan 2012).

Dhaka's suburbs surrounded with unequally distributed slums and squatter settlements. This is due to the high demand for land and land prices in the central zones. Since slum dwellers cannot afford to live in those regions, they move or are moving to peripheral city areas for cheaper land and settlement prices (Mahbub and Islam 1991; CUS 1993). Most of them are located in areas like Mirpur, Mohammadpur and Demra (BBS, 1999). A large number of slum dwellers were evicted from central parts of the Dhaka city by the urban development authorities which forced them to move to the peripheral areas (Hossain 2008).

A majority settlers work in low-paid jobs and mostly in the informal sectors (Hossain 2010). With a wide variation in occupational engagements between male and female slum dwellers, the former engage in eighty different types of occupations such as day laboring and rickshaw pulling, whereas the latter serve as maids and household chore workers (Hossain 2010; CUS 1983, 2006;

Amin 1991). Slum dwellers involved in the formal sector of the economy, and they are more privileged than their counterparts working in the informal sectors. Since most of the slums are temporary, they are made of weak infrastructural materials such as bamboo, tin, and wood. Over the decades, these slums mushroomed in both private and public lands. They are mostly located in low-lying flood prone areas, near railway tracts and risky canals, or in any privately owned lands (Hossain 2010). These make them easily exposed to eviction, natural and human made disasters. As the government does not consider these informal settlements as legal, urban services such as safe water, gas supply, electricity, hygienic toilet facilities and garbage disposal for the slum dwellers are highly restricted. Though in theory all urban population have equal access to public health and education facilities, in reality, slum dwellers have very limited access to both of those (Fariduddin and Khan 1996; Arnold 1997; Hossain 2001; World Bank 2007). In fact they are the most deprived group of people in the city. Both in primary and secondary education, their participation is very low.

In summary, Dhaka may have emerged as a fast growing metropolitan in the world, but the inequality and gap between the rich and the poor in the city are enormous and ever rising. The city faces gargantuan challenges in housing and in almost all aspects of its infrastructure that are supposed to serve the most basic needs of the city dwellers (Hossain 2010). The 11$^{th}$ goal of the Sustainable Development Goals (SDG) advocates for making cities and human settlements inclusive, safe, resilient and sustainable. With such challenges persist, massive tasks need to be undertaken in Dhaka city before it can achieve the 'equity and inclusivity' by 2030 underpinned by the 11$^{th}$ goal.

## Initiatives of the government of Bangladesh

The government of Bangladesh has agreed upon improving the urban housing and lands by undertaking a variety of policies and programs such as the draft National Urban Sector Policy, the Outline Perspective Plan, the sixth Five Year Plan. UN's Development Assistance Framework emphasizes improving the lives of urban poor specifically. In addition to developing organizations working on urban improvement schemes, number of urban development projects exist in Dhaka and other metropolitan areas, such as Urban Partnerships for Poverty Reduction, Urban Governance and Infrastructure Improvement Project and the Urban Primary Health Care Project.

## Role of NGOs and development partners

In addition to Government of Bangladesh (GOB), Non-Governmental Organizations (NGO) and development partners, including United Nations (UN) agencies and many bilateral and multilateral donors are active in the urban development activities in Bangladesh. During 1982-1985, the Department of Social Services undertook a United Nations Children's Fund (UNICEF)funded urban project, first of its kind in Bangladesh. The purpose of the project was to provide loan and health care to women, build day-care centres, set up tube wells and latrines in slums. However, the project faced many obstacles due to inexperience, frequent transfer of staffs and shortage of staffs (UNICEF 1988). Later on a number of different projects have been implemented by Local Government Body and Engineering Department (LGED) to provide the basic service to the city dwellers. These implemented projects focusing on partnership, women empowerment and participation.

Urban Governance and Infrastructure Improvement (Sector) project (UGIIP) by Asian Developmet Bank (ADB) initiated a performance-based allocation of funds. This strategy has proved to be an effective way to motivate the municipal authority and governance. For its second phase of the project, ADB partnered with GIZ[2] and KFW[3]. Recently it consulted with the Agence Francaise for the Greater Dhaka Sustainable Urban Transport Project. Along with UGIIP, World Bank participated in Municipal Services Project. It helped the government to establish strategies for urban transport project in Dhaka. Right now the World Bank is implementing the Clean Air and Sustainable Environment Project in Dhaka which focuses on traffic management and network restructuring in the centre of the city. Other development partners such as The Department for International Development of the United Kingdom (DFID), the United Nations Development Program (UNDP), and the United Nations Population Fund (UNFPA) work on areas like urban poverty and health care. The Japan International Cooperation Agency (JICA) has been supporting on solid waste management projects in Dhaka.

---

[2] GIZ= Deutsche Gesellschaft für Internationale Zusammenarbeit
[3] KFW= Kreditanstalt für Wiederaufbau ("Reconstruction Credit Institute")

A number of large and small NGOs are also involved in urban development initiatives particularly in the field of health, education, water, gender, sanitation, and hygiene promotion activities in the urban areas particularly in slums and squatter settlements. (Appendix I and II provide further details on the programs)

This paper, however, attempts to explore what types of relevant NGO programs and interventions are currently in place taking the case of Dhaka city. The paper critically seeks to identify the empirical evidence of 'equity and inclusivity' by landscaping the present programs, as well as previous programs of NGOs which were aligned with the former Millenium Development Goals.

## METHODS

A systematic review of urban programs and interventions implemented in Bangladesh in the last few decades has been carried out for this research. The review was based on program documentation available online including project briefs, baseline/end line reports, project documentation of interventions, and final evaluation reports. Activities undertaken as part of this review included: identification of urban development programs in Bangladesh with a focus on slums, poverty and social inclusivity, a collection of relevant programmatic information, and analysis of selected programs.

The review was primarily based on secondary literature collected through an online search of electronic databases, including Google, Google Scholar, JStor, Taylor & Francis Online etc. Search terms used included "urban development projects," "urban poverty reduction projects in Bangladesh," "slum and squatter improvement projects," "urban health services," and "urban education project." The search also included the names of the international and national NGOs engaged in urban development-related projects in Bangladesh.

The search was extended to the websites of government ministries, NGOs, and research organisations engaged in integrated urban development and management issues. These organisations include the Bangladesh Rural Advancement Committee (BRAC), CARE, Plan International, ActionAid, Dushtha Shasthya Kendra (DSK), Habitat for Humanity, Nari Uddug Kendra (NUK), UNDP, UNICEF, UNFPA, ADB and World Bank, and among others.

## RESULTS AND DISCUSSIONS

### A. THEMATIC FOCUS AREA

This review took the UN-HABITAT/OHCHR criterion of adequate housing to explain the thematic focus areas of the 20 programs and interventions identified in this review.[4]

The review identified that

- Approximately 11 of 20 programs focused on basic service provision to the slum dwellers, and at least five programs provide various types of financial incentives that help the inhabitants to access the basic needs.

---

[4] OHCHR and UN-HABITAT (2009), The right to adequate housing factsheet 21, 2009, p. 4. (Please see appendix 3)

- Seven programs have programmatic components on improving the habitability of the slum dwellings. At least five programs have components which are aimed to address the specific needs of disadvantaged and marginalized groups.

The analysis identified five trends in the thematic focus areas of reviewed projects:

1. **Land tenure in-security and dysfunctional national housing policy**

Land ownership and tenure insecurity are increasing problems in Dhaka. In 2008, only 30% of the total city's population held 80% of the lands, while the remaining 70% held only a meager portion of the land (Rashid 2009). The lack of ownership and affordable housing resulted in millions of people around the capital to find their own solution in various types of slums and unauthorised and informal settlements. People in these illegal settlements have been living a life of constant threat of eviction. They lack access to basic services such as electricity, clean water, sanitation and gas.

The National Housing Policy 2017 of Bangladesh recognises the basic needs of urban poor-- housing, shelter and food- which they are clearly deprived off. Through this document, government recognises that people residing in these settlements have been living a life of constant struggle and sufferings. The housing policy clearly acknowledges that evictions and forcible relocation of slum population are not a sustainable and humane option, and therefore should be avoided. Furthermore, the High Court through an order forbids forceful eviction. Despite the presence of such document and judicial restriction, slums and squatter settlements continue being demolished by successive governments without proper rehabilitation (COHRE and ACHR 2000; Nahiduzzaman 2006). They had ignored the continuous protests by local and international human rights organisations. During 1975-2005, around 135 slums and squatter settlements were demolished, and slum population had been evicted from their homes (Nahiduzzaman, 2006).

As the biggest service providers to poor population, NGOs found tenure insecurity and threat of eviction hindered their work in the urban slum settlements. They are reluctant to make any kind of capital investment there. Although, large NGOs such as BRAC, maintain education, health, nutrition, gender, and community empowerment programs in the slums and can cope with the threat of the eviction, many smaller NGOs cannot even afford to take such risks. For them, if a slum settlement becomes a subject to eviction, their capital investment is at stake, along with their investment in staffing time and training (Rashid 2009).

For instance, the Agargaon slum was one of the largest informal settlements in Bangladesh until it was demolished. The formation of the slums took place in the mid-80s and mushroomed for more than 20 years on government land. Then in 2004, the government decided to demolish the slums in order to establish government institutions. It became the largest eviction in Dhaka city which affected the lives of around 40,000 people who were mostly poor and vulnerable (COHRE and ACHR 2000). A number of NGOs were implementing different projects in that slum settlement. They lost a vast amount of capital invested there. Plan International-- a prominent INGO was running water and sanitation, non-formal education and health programs at the slum. Other NGOs like Proshika and Dushtha Shasthya Kendra (DSK) operating microcredit programs. But, Proshika and DSK's project loss were significantly smaller than the larger and more infrastructural programs implemented by Plan International. As a consequence, most NGOs lost their faith, and barred themselves in implementing projects in such settlements.

2. **Action for safe and affordable housing**

Many attempts have been taken to rehabilitate the poor population and build housing for them. But, institutional rigidity and political unwillingness from the government and other parties stymied the progress. During its first five year plan (1973-1978), the government decided to build housing for low-income entities.

After evicting 200,000 slum residents in 1975, government took the initiative to resettle the residents in various parts of Dhaka (COHRE and ACHR 2000). One of such sites was a low undeveloped land of Bhasantek, Mirpur where 3,044 families were rehabilitated. These families were again transferred to another location. UN Center for Housing, Building and Planning provided 2,300 families shelter with basic human utilities and community facilities (Shafi 2005). The project was known as Kalshi resettlement project. Each family was given a 475 sq ft of plot (Rashid 2009). However, most of these resettled inhabitants did not have financial solvency. Therefore, they often sold their housing for meagre amount of cash and moved back to other slum settlements. A number of microcredit projects introduced a strategy where a group of people act as guarantors for each other, ensuring the land is not misused or handed over.

The Ministry of Land committee recommended a comprehensive plan in 1989 to reduce the slum burdens from the Dhaka metropolitan areas. But due to institutional rigidity, this plan never came out to see the light. Again in 1990, the then government had taken town-housing projects like- Ghore Phera, Asrayan, Adarsha Gram Prokolpo, which were expected to encourage people to return to their villages (Shafi, 2005). The assumption was that with proper incentives people living in slums will return to their respective villages with an enormous social and economic network. However, these projects failed due to lack of employment opportunities in the villages and lack of proper monitoring as well (Rashid 2009).

A number of government authorities like Ministry of Public Works and Housing (MPWH), Local Government Body and Engineering Department (LGED) of Bangladesh have taken many projects over time to address the urban housing problems in Dhaka and its surroundings. In 1999, a private company named- North-South Development was appointed to build housing for the poor (Rashid 2009). Indeed, the project started in that year, but never really saw its completion. Though different authorities have introduced different projects, very little progress was made in implementing National Housing Policy property rights. Policies to improve the lives of slum people are still inadequate and uncoordinated. The number of failed projects prove that it is still the biggest challenge in urban Bangladesh.

NGOs, on the other hand, are reluctant to involve in urban housing projects. The main reasons are lack of urban lands and high recovery costs as well as the threat of eviction as mentioned in the earlier section (Shafi, 2005). Nevertheless, some small NGOs took smaller housing development initiatives. Nari Uddyog Kendra (NUK) are working on urban housing for women. Female garment workers constitute a large group in the slum settlements of greater Dhaka. They basically settle around the industrial areas. Local goons control most of these slum settlements, and the female workers become victim of these goons. NUK took a project to respond to the housing needs of these women. It introduced hostel services for these women with 600-bed capacity both in Dhaka and Mymensingh (Rashid 2009). Recently, similar hostel facility has been introduced by BRAC in collaboration with government.

### 3. Exclusion from the very basic rights by virtue of residence

Constitution of Bangladesh, article 15, clearly states the basic responsibilities of the state to its citizens. It states that there are five basic necessities- food, clothing, shelter, education and medical care. The constitution compels the state to ensure nutrition and public health along with equality to all its citizens. But the irony is that slum people has to fight for the very basic constitutional rights in their every sphere of life.

As people need an official address to access the basic rights provided by the state, most slum population get rejected due to the illegal nature of the establishments. A survey of 9,048 slums in six major cities in Bangladesh found that more than half of the slum population does not even have a fixed and hygienic garbage disposal facility (Islam et al. 2006) The few who receive the basic services have to pay very high price. Most of the time these poor people have to pay high prices to the local goons and political leaders for water and other services (Rashid S. , 2005). Local goons bribe the municipal officers to get illegal connections. They charge high prices (Tk 5 per bucket, or Tk. 1-2 per bath) for water from those sources (Rashid 2009). In places where water is free, slums dwellers have to stand in large ques to access that water. For example, in Phulbari Slum, women and girls have to stand on the line for hours and water is only accessible for half an hour every day (Rashid 2005). Electricity connection is also under the local goons' control, and slum people have to pay exorbitant prices for that. Coalition for the Urban Poor (CUP) states that the price paid by the slum dwellers are three times higher than what regular citizens pay for their electricity consumption (Islam et al. 1997).

Sanitary latrine is almost inexistent in slums in Dhaka. 70% of the establishments had no access to sanitary latrines (Rashid 2009). In most slums, latrines are shared with other families. In half of the cases around 30 people share a single latrine (Islam et al. 2006). In Korali Bastee (the biggest slum in Dhaka), where more than 12,000 families reside, there is no toilet or hearth clinic (South Asia Region, World Bank 2007). The drainage system and sanitation is very poor in slums. A survey of 9,048 households in slum reveal that about 26.5% of the slum settlements experience full-fledged flood in every rainy season (Islam, 2006).

DSK has implemented a policy to ensure clean water supply to the urban poor of Dhaka especially those living in slums and squatter settlements (Ahmed 2003). The success of its program has encouraged many donor organizations like- UNDP, World Bank Water and Sanitation Program, Swiss Agency for Development and Cooperation and Water Aid to work on similar programs in other areas. Lessons from its program also motivated Dhaka Water Supply Policy to introduce community participation to raise awareness among stakeholders and to monitor government policy and plans.

4. **Accessibility for women, children and people with disabilities**

The financial situation of urban slum people can be improved through scaling up microfinance programs to urban areas. With this idea in mind, Shakti Foundation for Disadvantaged Women initiated a project based on Dhaka in 1992 (Rashid 2009). It financed poor women living in slums in Dhaka. Around $76 million was distributed for trade, manufacturing, processing and services. It provided loans ranging from $57-$71 for income-generating schemes and ideas (Islam et al. 2006). In Dhaka, it distributed loan to over 105,000 people (Rashid 2009). One-third of the borrowers have managed to cross poverty line and around 10% have managed to cross the lower middle-income threshold (Islam et al. 2006).

Security of women and girls are a burning issue. ActionAid Bangladesh implemented Safe Cities for Women Campaign with the aim to increase awareness against sexual violence in public spaces among women and girls. There are many similar NGO initiatives.

5. **Poor communication and transportation: a barrier to promoting equity and inclusivity**

One of the key barriers to slum improvement and dwelling upgrading is poor transportation and connectivity. Most of the slums do not have adequate roads for easy movement. For instance, most people in Korail slum has to use boats to come to the main city area as there is only one undeveloped road connected. Not only the slum settlements, but poor road communication is the picture of entire Dhaka city. One of the key components of achieving 'equity and inclusivity' in a Metropolitan is to have a mass rapid transit system including well-organized road and railway connection.

In major cities such as Dhaka, private sector transportation plays a key role. But as the public monitoring system is not strong enough, they do not comply with the urban transportation rules and regulations. It creates congestions and traffic all around the cities. Accidents and injuries are quite frequent. Moreover, the mixture of motorised and non-motorized vehicles in the same street causes unbearable traffic congestion in the Capital city. There is no separate lane for bicycles and rickshaws or street for pedestrians. Indeed, a 37 km long railroad passes through the city, but it's contribution is limited due to effective policy mechanism (BUF, 2011). The rail crossings are weakly maintained and they are dangerously prone to accidents. There are 100s of open market on the streets and around 3,000 roadside shopping centres with little or no parking facilities (BUF, 2011).

### B. PROGRAMMING APPROACHES

#### i. Awareness raising and knowledge building approach

This review found awareness raising and knowledge building to be the most common urban improvement strategy to address the social exclusion in the slums and squatter settlements in Dhaka city and its' fringes. Some programs (such as UPPR and EMPOWER) were designed to create networks of government, non-government and private sector urban service providers who provide access to quality, affordable services to clients living in poverty. Some programs (such as Building Resilience of Urban Slum Settlements, and Urban Wash programme) sought to send their message through billboard advertisement, dramas, theatre for development approach to ensure the water, sanitation, and hygiene management in their respective intervention areas.

Security for women particularly for working women is a burning issue in Bangladesh. Project such as SHE CAN has embarked on campaigns in the cities leading to greater respect for women and girls' rights in every aspect of urban life. The program sensitises women and girls about services such as legal aid, victim support and other judicial supports. Bangladesh is one of the top readymade garments exporters. More than thousand garments factories are operating within and the fringes of the capital city. Most of the workers in these factories are women. These factories frequently keep overnight shifts. This makes the women workers to get back to their house at late night. Some programs (such as Garment Factory & Workers Support Program) tirelessly lobbying for the provision of transport for women garment workers, particularly at night.

**Advocacy activities**

Another approach is the advocacy activities carried out to formulate prescriptive guidelines for future sustainable urban development. BRAC Institution of Governance and Development (BIGD) has partnered with Cities Alliance and UN Office for Project Services (UNOPS). They will work with local governments, city stakeholders and development partners. Their project plans to conduct research and advocacy to produce global knowledge. The project aims to facilitate policy dialogues. The project also supports city-level diagnostics thus providing policy recommendations to respond to the challenges of the inequitable economic growth in the Dhaka city.

**ii.    Community mobilization**

The community mobilization has also been extensively used in urban programming, typically in conjunction with the awareness raising and knowledge building models. Elements of community mobilisation were found in eight programs out of the 20 programs reviewed. Programs (such as EMPOWER and UPPR) are designed to engage poor urban communities and slum residents in participatory group activities in developing community action plans. The process is inclusive to identifying key problems, finding sustainable solutions and putting them in action. One of the urban improvement schemes is disaster preparedness. Slum and squatter settlements are poorly constructed and critically exposed to natural disasters such as storm, flood and erosion. Program such as BRUP is working to develop self-sustainable model of disaster resilience. For this purpose, they are forming empowered women group, urban volunteers, and better risk-informed community. BRUP also liaised with government bodies to strengthen institutions for providing supports at doorstep to the vulnerable people.

iii. **Service delivery approach**

   a. **WASH and Waste Management**

A number of NGOs are active in Bangladesh in water, sanitation and hygiene (WASH) promotional activities. Practical Action, WaterAid, Plan International, BRAC have large projects on this across the country. However, two major programs (Urban Wash programme and DSK model) have extensively worked on the WASH issues in the informal urban settlements in Dhaka. Therefore, the review included these two programs exclusively.

Dhaka WASA has a policy of not providing water services to the houses who do not possess legal permits. Almost all of the slum dwellers become subject to this restriction. In 1992, DSK agreed with the government that DSK will act as the guarantor for the security deposit and payments of bills of the people who lived in slum settlements (Ahmed 2003, Rashid 2009). WASA opened water points which till date considered as a great model of success. Another prominent NGO WaterAid is working on WASH issues in Bangladesh since 1986. In 2011, WaterAid scaled up its urban program and introduced the 'Urban WASH programme'. The program services provided include:

- the installation of tube wells;
- community/cluster latrines with septic tanks;

- the construction of sanitation blocks combining water points, bathing stalls and hygienic latrines;
- the construction of footpaths;
- household water-seal, pit latrines;
- drainage improvements;
- solid waste management; and
- hygiene education.

Some programs (such as 'Building Resilience of Urban Slum Settlements: A Multi-Sectoral Approach to Capacity Building- Habitat for Humanity' and 'SWEEP') extensively working on solid waste and sludge management. However, the Habitat for Humanity program has carried out pilot activities on housing improvement and on WASH initiatives. They are planning to scale up their activities in the coming years.

b. **Microcredit loan schemes**

BRAC, ASHA, Proshika, and Grameen are the pioneer organisations who offer micro-credit loans. Two programs in this review are identified (Micro Finance program and Urban Poor Development Program) which offers Micro-credit loans to the poor urban community particularly women. These loans are offered to initiate various income-generating activities.

c. **Education and Health Care**

Several NGOs are engaged in education and health care programs in the slum areas. Programs such as BRAC urban school program have established hundreds of schools which provides primary education to the slum children. Plan International also maintains same education programs.

Two projects (Manoshi and CSN) of BRAC provides health care and special need services to the slum dwellers.

## CONCLUSION

Dhaka is an urban agglomeration which is growing hastily. The slow pace of development in the housing sector, infrastructure, and transportation in the city cannot commensurate with its rapid population growth. This is an immense barrier to the socio-economic progress of Bangladesh. People who failed to secure a decent and adequate housing in the city has found their own solutions of living in the informal settlements such as slums and squatters. Government considers these settlements as illegal and is reluctant to provide the most basic services- water, sanitation, education, healthcare, gas, electricity and also legal services there. To circumvent these obstacles, NGOs with the support of donors started intervening in the slums. This review shows that most of the NGO programs have invested their time and money in water, sanitation, hygiene improvement and solid waste management programs. Others invested on education, health care, gender and microfinance. Large NGOs have carefully tailored their interventions in the slums because of the threat of eviction, and slum relocation. But smaller NGOs are afraid to invest time, money and manpower, because if any day the slums are evicted, all of their assets and investment will be wasted. Nevertheless, innovative programs such as DSKs water model, BRAC's urban slum school program, or microfinance programs of BRAC, Proshika, Shakti Foundation have contributed to a great extent to improve the lives of millions of slum dwellers. Care, WaterAid, Habitat for Humanity and similar likeminded NGOs have also changed their lives through their dynamic project interventions. Despite their great efforts and achievements, some challenges remain. The review identified that there are duplication of efforts, as multiple NGOs have interventions in slums which are similar in nature. Lack of coordination among themselves and absence of an ideal model are often cancelling out their achievements. Furthermore, frequent eviction and slum relocation threatens the intended project outcomes. In this regards, a citywide in-depth mapping and review of their programs is required to come up with an innovative and ideal model. This innovative model will bring the Government, NGOs, and development partners in the same portfolio, and will lead to developing a concerted and well-coordinated action plan.

Hossain, S. (2001). Research on slums and squatters in Bangladesh: a critical review,. *Social Science Review, 18*(2), 67–76.

Hossain, S. (2008). Rapid urban growth and poverty in Dhaka City. *Bangladesh e-Journal of Sociology, 5*(1).

Hossain, S. (2010). *Urban Poverty in Bangladesh: Slum Communities, Migration and Social Integration* (Vol. 3). I.B.Tauris.

Islam, N. (2006). In R. Brian, & T. Kanaley, *Urbanization and sustainability in Asia* (pp. 43-70). Manila: Asian Development Bank.

Islam, N., Huda, N., Narayan, F., & Rana, P. (1997). *Addressing the urban poverty agenda in Bangladesh: critical issues and the 1995 survey findings.* Dhaka: University Press Ltd.

Islam, N., Mahbub, A., Nazem, N. I., Lance, P., & Angeles, G. (2006). *Slums of urban Bangladesh: mapping and census.* Dhaka: Centre for Urban Studies.

Islam, N., Mahbub, A., Nazem, N., Lance, P., & Angeles, G. (2006). *Slums of urban Bangladesh: mapping and census 2005.* Dhaka: Centre for Urban Studies.

Jahan, S. (2012). Managing the Urban Transition in Bangladesh. In S. Ahmed, *Leading Issues in Bangladesh Development .* Dhaka: University Press Limited .

Mahbub, A., & Islam, N. (1991). The growth of slums in Dhaka City: a spatiotemporal analysis. In S. U. Ahmed, *Dhaka Past Present Future* (pp. 508–21). Dhaka: Asiatic Society of Bangladesh.

Mohit, M. A. (2012). Bastee Settlements of Dhaka City, Bangladesh: A Review of Policy Approaches and Challenges Ahead. *Procedia - Social and Behavioral Sciences, 36*, 611- 622.

Nahiduzzaman, M. (2006). *Housing the urban poor: planning, business and politics: a case study of Duaripara slum, Dhaka city, Bangladesh.* Trondheim: Department of Geography, Faculty of Social Sciences and Technology Management, Norwegian University of Science and Technology.

Nari Uddug Kendra (NUK). (2018). *Garment Factory & Workers Support Program .* Retrieved January 1, 2018, from http://www.nuk-bd.org/: http://www.nuk-bd.org/garment_support.php

OHCHR and UN-HABITAT. (2009). *The right to adequate housing factsheet 21.*

Plan International Bangladesh. (2018). *Developing a model of Inclusive Education in Plan Bangladesh.* Retrieved February 3, 2018, from https://plan-international.org/: https://plan-international.org/developing-model-inclusive-education-plan-bangladesh

PPRC. (2009). *Urban Resident Survey.* Power and Particpation Research Centre .

**APPENDIX I: URBAN DEVELOPMENT PROGRAMS IN BANGLADESH 1975- PRESENT**

| Sl. | Program | Year | Donor | Implementing Organization | Program Reach & Location | Project Theme | Approach & Project Description | Target Group | Source |
|---|---|---|---|---|---|---|---|---|---|
| 1 | **UPPR** Urban Partnerships for Poverty Reduction | 2000-present | UNDP, DFID | Rural Development and Co-operatives, UNHABITAT, LGED ministry-GOB | 23 cities and towns including Dhaka and greater Dhaka region (Naraynganj, Tongi, Gazipur) | Basic Services (Access to water, sanitation, electricity, waste management) | **Community Mobilization** Mobilize Urban poor communities to form representative and inclusive groups and prepare community action plans. **Service Provision** Support the communities to meet demands for water supply, sanitation, drainage, electricity and public lighting, waste management, Provide road access and community facilities through participatory processes. **Awareness Raising and Advocacy** Facilitate urban-poor policy dialogue through networking of towns, association of elected representatives, LG officials and community leaders. Development and implementation of communications strategy for programme information sharing, advocacy and policy dialogue. | Urban poor and extremely poor people, especially women and children. | Please follow the link |

| Sl. | Program | Year | Donor | Implementing Organization | Program Reach & Location | Project Theme | Approach & Project Description | Target Group | Source |
|---|---|---|---|---|---|---|---|---|---|
| 2 | EMPOWER Engaging Multi-sectoral partners for Creating Opportunities, Improving Wellbeing and Realizing Rights of the Urban Poor | 2015-present | _ | BRAC | Seven city corporations (including Dhaka and Narayanganj) | Availability of Basic Services | **Community Mobilization** Help slum residents form groups to make their voices heard. Assist women and young people to take the lead in identifying key problems, finding sustainable solutions and putting them in action. **Awareness Raising and Advocacy** Create a network of government, non-government and private sector urban service providers who provide access to quality, affordable services to clients living in poverty. Using this network, a referral system is intended to develop where clients can learn about, demand and use the best available services within their reach. Work with policy makers and influencers to improve urban policies and implement them properly. | 500000 people from 150 slums of 7city corporations | Please follow the link |

| Sl. | Program | Year | Donor | Implementing Organization | Program Reach & Location | Project Theme | Approach & Project Description | Target Group | Source |
|---|---|---|---|---|---|---|---|---|---|
| 3 | Community Fire Prevention Project | 2015 | _ | BRAC | Dhaka | Habitability | **Awareness Raising** Promote changes in attitudes towards fire safety. This is intended to be implemented through two initiatives- partnering with the Bangladesh Fire Service and Civil Defense, to help communities learn about ways to prevent fire, collaborate with the community and the Fire Service to help establish simple firefighting mechanisms for when fires break out. | Slum people in Dhaka | Please follow the link |
| 4 | Urban Innovation Challenge | 2015 | Australian Aid, UKaid | BRAC | Dhaka, Chittagong, Sylhet | Location, Availability of Basic Services | **Awareness Raising and Advocacy** Choose a small team through competition of early-stage organization (for-profit or non-profit) who have solutions or can design innovative solutions with potential to benefit over 1 million people living in any one of Bangladesh's cities | Small team of early-stage organization (for-profit or non-profit) | Please follow the link |

| Sl. | Program | Year | Donor | Implementing Organization | Program Reach & Location | Project Theme | Approach & Project Description | Target Group | Source |
|---|---|---|---|---|---|---|---|---|---|
| 5 | BRUP Building Resilience of the Urban Poor | 2014-2017 | C & A Foundation | Care-Bangladesh | Gazipur District (Greater Dhaka) | Habitability | **Community Mobilization** To increase resilience of urban communities and institutions so that they can prepare for, mitigate, respond to, and recover from socks and stresses.<br><br>To develop a 'Self Sustainable Model of Resilience 'which will consist of empowered women group, urban volunteers, better risk informed community and strengthened institutions for providing supports at door step to the vulnerable people. | 8000 people living in informal and slum settlements | Please follow the link |
| 6 | SHAHAR Supporting Household Activities for Health, Assets and Revenue | 1999-2004 | USAID | Care-Bangladesh, IFPRI | Jessore and Tongi (Greater Dhaka) | Affordability, Availability of Basic Services | **Community Mobilization** Strengthen community, NGO, pourashava, and GOB partners' capacity to maintain sustainable urban programs.<br><br>**Service Provision** To increase incomes and improve unhygienic environments and sanitation conditions in poor urban communities, including slums, in major selected secondary cities in Bangladesh | 271 households in the municipal areas of Jessore and Tongi | Please follow the link |

| Sl. | Program | Year | Donor | Implementing Organization | Program Reach & Location | Project Theme | Approach & Project Description | Target Group | Source |
|---|---|---|---|---|---|---|---|---|---|
| 7 | Building Resilience of Urban Slum Settlements: A Multi-Sectoral Approach to Capacity Building | 2012-2013 | AusAID – ANCP Innovations Fund | Habitat for Humanity Bangladesh | Talab Camp in Mirpur, Dhaka | Habitability, Availability of basic services | **Community Mobilization** Take day-long cleaning event was undertaken in Talab Camp with the community and HFHB volunteers. **Awareness Raising** Use of Bill Boards to send message "My neighbourhood, my house, let's clean them regularly" to the beneficiaries **Service Provision** Solid waste management, housing improvement, WASH initiatives | An urban slum community of 650 households known as Talab Camp in Mirpur in the north-western part of Dhaka | Please follow the link |
| 8 | **Urban Wash programme** | 2011-till date | SIDA, WaterAid | WaterAid | Dhaka, Chittagong and Khulna | Availability of basic services | **Awareness Raising** Organize drama, theatre, and other cultural activities at community level engaging mass people as well as other audiences to act towards the cause of safe water, hygiene and sanitation. **Service Provision** Solid waste management, WASH initiatives | Total of almost 340,000 people are expected to benefit with domestic WASH and nearly 1.7 million commuters and members through the use of mobile and static public toilets | Please follow the link |

| Sl. | Program | Year | Donor | Implementing Organization | Program Reach & Location | Project Theme | Approach & Project Description | Target Group | Source |
|---|---|---|---|---|---|---|---|---|---|
| 9 | **SHE CAN** | 2014- till date | UKAID | ActionAid | Narayanganj (Greater Dhaka), Chittagong, Sylhet, Rajshahi, Barisal, Rangpur and Khulna in Bangladesh | Accessibility | **Awareness Raising** Ensure women and girls are aware of their rights and know how to take actions to enhance their safety and access to justice (legal aid, victim support etc.). Campaign in the cities in order to achieve greater respect for women and girls' rights and public support for the Safe Cities campaign from women, girls, men, boys and duty-bearers. **Community Mobilization** Mobilize women and girls' networks and coalitions mobilized and supported to actively lead local and national solidarity movements to demand an end to violence against women. Duty-bearers, employers and public and private sector services in the cities are more responsive to women and girls' rights to Safe Cities. | Women, girls, men, boys and duty-bearers | Please follow the link |
| 10 | **Promoting Equitable Economic** | 2016-present | DFID, CLGF, Ford Foundatio | Cities Alliance, UN Office for Project Services (UNOPS), BIGD | Narayanganj and Sylhet | Affordability, Habitability, Availability of basic | **Awareness Raising and Advocacy** Cities Alliance and BIGD aims to work with local | Local authorities and developm | Please follow the link |

| Sl. | Program | Year | Donor | Implementing Organization | Program Reach & Location | Project Theme | Approach & Project Description | Target Group | Source |
|---|---|---|---|---|---|---|---|---|---|
| | Growth in Cities | | n, UNCDF), UN-Habitat, WIEGO, the World Bank | | | services, Accessibility | governments, city stakeholders and development partners will conduct research and advocacy to produce global knowledge, facilitate policy dialogues and support city-level diagnostics and policy recommendations to respond to the challenges of the inequitable economic growth in cities. | ent partners | |
| 11 | Garment Factory & Workers Support Program | 1992-present | Fair Wear Foundation | Nari Uddog Kendra (NUK) | Dhaka and Mymensingh | Habitability, Availability of basic services, Accessibility | **Awareness Raising and Advocacy** Lobbies for Quality Standards which exceed the requirements of the basic social compliance audit. Lobbies for provision of transport for women garment workers, particularly at night.<br><br>**Service Provision** NUK was a pioneer in providing women worker hostels which are widely recognised as models for clean, safe, secure and low cost dormitory accommodation. | Female garments workers living in the fringe of Dhaka city | Please follow the link |

| Sl. | Program | Year | Donor | Implementing Organization | Program Reach & Location | Project Theme | Approach & Project Description | Target Group | Source |
|---|---|---|---|---|---|---|---|---|---|
| 12 | **DSK Model** | 1992-present | - | Dushtha Shasthya Kendra (DSK) | Dhaka | Availability of basic services | **Awareness Raising and Advocacy** The DSK model prompted the water authority to allow communities to apply for water connections on their own behalf without the need for a guarantor. **Service Provision** Eighty-eight water-points have been established in 70 slum-settlement areas since 1996, About 12 water-points have been paid for and handed over to user-groups. The water authority is also cooperating in the replication of the project in 110 community-managed water systems | 450,000 | Please follow the link |
| 13 | **Micro Finance program** | 1992-present | - | Shakti Foundation | Dhaka, and other districts | Affordability | **Service Provision** Provide Microcredit loans range from BDT 4,000 to BDT 50,000 to support income-generating activities | Covered 325 branches across 53 districts and supported over 2 lakh and 12 thousand women till date. | Please follow the link |

| Sl. | Program | Year | Donor | Implementing Organization | Program Reach & Location | Project Theme | Approach & Project Description | Target Group | Source |
|---|---|---|---|---|---|---|---|---|---|
| 14 | **Urban Poor Development Program** | 1975-present | - | Proshika | Dhaka, Chittagong and Khulna | Affordability | **Service Provision** BDT 6.92 million has been disbursed to 578,027 borrowers against 92,740 projects to promote different employment and income-generating activities of the urban poor | Nearly 25,275 groups have so far been formed in the slum areas. | Please follow the link |
| 15 | **SWEEP** | 2015-present | DFID | WSUP | Dhaka | Habitability and Availability of Basic Services | **Service Provision** Solid Waste management- cleaning septic tanks in residents of Dhaka. Provide door to door service. | 102,408 residents in Dhaka | Please follow the link |
| 16 | **Manoshi** | 2009-present | - | BRAC, ClickDiagnostics | Dhaka, Korail Slum area | Availability of Basic Services, Location, Habitability | **Service Provision** The Shasthya Shebikas (SS) and Shasthya Kormis (SK) provide antenatal and postnatal care, essential newborn care (ENC) and child health care. Connects community with health facilities via an innovative mobile phone based referral system known as 'ClickDiagnostics'. **Awareness Raising** Through behaviour change communication interventions Shasthya Kormis to motivate, educate and prepare expectant mothers for childbirth, and other reproductive issues. | 223,487 people | Please follow the link |
| 17 | **Dabi Microfinance project** | 1974-present | - | BRAC | Dhaka slums and all districts | Affordability | **Service Provision** Provides loan to women. | 1,900 members and 770 | Please follow the link |

| Sl. | Program | Year | Donor | Implementing Organization | Program Reach & Location | Project Theme | Approach & Project Description | Target Group | Source |
|---|---|---|---|---|---|---|---|---|---|
| | | | | | | | The average loan size is BDT 8,227 (USD 121). | borrowers in Korail slum areas, and 8.02 million | |
| 18 | **CSN Children with special needs** | 2003-present | - | BRAC | Dhaka and surrounding areas | Accessibility | **Service Provision** Provides corrective surgeries, along with devices like wheelchairs, crutches, hearing aids and glasses; and even builds ramps to make classrooms more accessible to disabled.<br><br>Printed textbooks in Braille to support the children who are visually impaired, and has also trained the staff on it. | 1,78,921 children | Please follow the link |
| 19 | **Creating a role model of Inclusive Education** | 2016-present | - | Plan International | Dhaka, Nilphamari, Lalmonirhat, Barguna, Bhola | Accessibility | **Community Mobilization and Awareness Raising** Build rapport for inclusive education at the community, school, and governance level by raising awareness, building capacity and skills, and changing the attitude of the existing system | 20,000 students in 50 primary schools | Please follow the link |

| Sl. | Program | Year | Donor | Implementing Organization | Program Reach & Location | Project Theme | Approach & Project Description | Target Group | Source |
|---|---|---|---|---|---|---|---|---|---|
| 20 | **BEP BRAC Education Programme- BRAC Urban Slums Schools** | 1985-present | EAC | BRAC | Dhaka, Sylhet, Chittagona, Rajshahi, Khulna, Barisal, Rangpur, Jessore, Mymensingh, Commilla, Gajipur, and Norsingdi | Location, Accessibility | **Service Provision** 2,000 one-room schools to urban slums or areas adjacent to the slums<br><br>**Community Mobilization** More than 10,000 parent committee members are being trained on school monitoring and children's attendance monitoring in an effort to support both the teaching and learning outcomes in local schools. | 62,000 children | Please follow the link |

*Source: Brac.net; http://www.carebangladesh.org/; http://www.habitatbangladesh.org/; https://plan-international.org/bangladesh ;*

*http://www.wsup.com/programme/where-we-work/bangladesh/; www.proshika.org/; www.sfdw.org/; www.dskbangladesh.org/; nuk-bd.org/; bigd.bracu.ac.bd/;*

*UNDP Bangladesh; www.wateraid.org/bd; www.actionaid.org/bangladesh.*

## ANNEXURE II: PROJECTS BY MAJOR DEVELOPMENT PARTNERS IN THE URBAN SECTOR

| Sl. | Name of the project | Development Partner | Project Duration | Project Theme | Project Cost (US$ m) |
|---|---|---|---|---|---|
| 1 | Slum Improvement Project | UNICEF | 1985-88 | Slum upgrading | 0.10 |

| Sl. | Name of the project | Development Partner | Project Duration | Project Theme | Project Cost (US$ m) |
|---|---|---|---|---|---|
| 2 | Slum Improvement Project II | UNICEF | 1988-96 | Slum upgrading | 4.60 |
| 3 | Secondary Town Infrastructure Development Project I (slum component) | ADB | 1992-97 | Slum upgrading | 0.62 |
| 4 | Secondary Town Infrastructure Development Project II (slum component) | ADB | 1996-01 | Slum upgrading | 1.28 |
| 5 | Secondary Towns Integrated Flood Protection Project II | ADB | 1992-98 | Flood protection | 0.61 |
| 6 | Urban Basic Service Delivery Project | UNICEF | 1996-01 | Integrated urban development | 5.8 |
| 7 | Community Empowerment for Urban | UNDP | 1996-01 | Urban Poverty | 10 |

| Sl. | Name of the project | Development Partner | Project Duration | Project Theme | Project Cost (US$ m) |
|---|---|---|---|---|---|
| | Poverty Alleviation | | | | |
| 8 | Municipal Services Project (slum component) | World Bank | 1995-00 | Integrated urban development | …. |
| 9 | Urban Poverty Reduction Project | ADB | 1998-02 | Urban Poverty | …. |
| 10 | Local Partnerships for Urban Poverty Alleviation Project | UNCHS/UNDP | 2000-07 | Urban Poverty | …. |
| 11 | Urban Partnership for Poverty Reduction | UNCHS/UNDP | 2008-15 | Urban Poverty | 120 |

| Sl. | Name of the project | Development Partner | Project Duration | Project Theme | Project Cost (US$ m) |
|---|---|---|---|---|---|
| 12 | Preparing the Greater Dhaka Sustainable Urban Transport Project (project preparatory technical assistance) | ADB | 2010–2011 | Urban transport | 1.00 |
| 13 | Dhaka Urban Transport Project | World Bank | 1995–2005 | Urban transport | 177.00 |
| 14 | Clean Air and Sustainable Environment Project | World Bank | 2009–2013 | Urban transport | 77.54 |
| 15 | Dhaka Urban Transport Network Development Studies | JICA | 2009–2010 | Urban transport | |
| 16 | City Region Development Project | ADB | 2010–2017 | Integrated urban development | 170.00 |
| 17 | Urban Governance and Infrastructure Improvement (Sector) Project | ADB | 2003–2010 | Integrated urban development | 60.00 |

| Sl. | Name of the project | Development Partner | Project Duration | Project Theme | Project Cost (US$ m) |
|---|---|---|---|---|---|
| 18 | Second Urban Governance and Infrastructure Improvement (Sector) Project | ADB/KfW/GIZ | 2008–2014 | Integrated urban development | 167.50 |
| 19 | Municipal Service Project | World Bank | 1999–2011 | Integrated urban development | 138.60 |
| 20 | Municipal Service Project (additional financing) | World Bank | 2008–2011 | Integrated urban development | 25.00 |
| 21 | Secondary Towns Water Supply and Sanitation Projects | ADB | 2006–2013 | Urban water supply and sanitation | 41.00 |
| 22 | Dhaka Water Supply Sector Development Project | ADB | 2007–2014 | Urban water supply and sanitation | 50.00 |
| 23 | Dhaka Water Supply Sector Development Program | ADB | 2007–2014 | Urban water supply and sanitation | 50.00 |

| Sl. | Name of the project | Development Partner | Project Duration | Project Theme | Project Cost (US$ m) |
|---|---|---|---|---|---|
| 24 | Dhaka Water Supply and Sanitation Project | World Bank | 2008–2017 | Urban water supply and sanitation | 167.50 |
| 25 | Water Supply and Sanitation Sector Program Support Phase I | Danida | 2006–2010 | Urban water supply and sanitation | 60.82 |
| 26 | Environmental Sanitation, Hygiene, and Water Supply Project in Slum Areas | UNICEF | 1997–2005 | Urban water supply and sanitation | 2.40 |
| 27 | Urban Public and Environmental Health Sector Development Program | ADB | 2009–2016 | Urban environment | 80.00 |
| 28 | Clean Air and Sustainable Environment Project | World Bank | 2009–2014 | Urban environment | 71.25 |
| 29 | Strengthening Solid Waste Management in Dhaka City | JICA | 2007–2011 | Urban environment | |

| Sl. | Name of the project | Development Partner | Project Duration | Project Theme | Project Cost (US$ m) |
|---|---|---|---|---|---|
| 30 | Urban Partnerships for Poverty Reduction Program | DFID/UNDP | 2007–2015 | Urban poverty | 120.00 |
| 31 | Urban Primary Health Care Sector Development Program | ADB | 2009–2017 | Urban health care | 80.00 |
| 32 | Second Urban Primary Health Care Project | ADB/Sida/DFID/UNFPA | 2005–2012 | Urban health care | 90.00 |
| 33 | Secondary Towns Integrated Flood Protection Project | ADB | 2006–2010 | Flood protection | 80.00 |
| 34 | Bangladesh Urban Resilience Project | World Bank | 2015-2020 | Integrated urban development | 182.00 |
| 35 | Good Governance in Urban Areas Programme | GIZ | 2012-2014 | Integrated urban development | …. |
| 36 | Pro-poor Slum Integration Project | World Bank | 2016-present | Slum upgrading | …. |

*Source: Rahman, Mohammed Mahbubur.2008 "Sustainability of Slum Improvement Program in Bangladesh: An Approach of Capacity Building, Community Participation and Empowerment." Journal of Bangladesh Institute of Planners ISSN 2075: 9363; Asian Development Bank*

*…. = not available, ADB = Asian Development Bank, DFID = Department for International Development of the United Kingdom, GIZ = Deutsche Gesellschaft für Internationale Zusammenarbeit, JICA = Japan International Cooperation Agency, Sida = Swedish International Development Cooperation Agency, UNDP = United Nations Development Programme, UNFPA = United National Population Fund, UNICEF = United Nations Children's Fund.*